\documentclass[%
 reprint,
 amsmath,amssymb,
 aps,
pre,
floatfix,
]{revtex4-1}

\usepackage{graphicx}
\usepackage{dcolumn}
\usepackage{bm}
\usepackage{epstopdf}
\usepackage{siunitx}
\usepackage{gensymb}
\usepackage{float} 

\begin{document}


\title{Time-domain imaging of curling modes in a confined magnetic vortex and a micromagnetic study exploring the role of spiral spin waves emitted by the core}

\author{D. Osuna Ruiz\textsuperscript{1}} 
\email{do278@exeter.ac.uk}
\author{P. S. Keatley\textsuperscript{1}}
\author{J. R. Childress\textsuperscript{2}}
\author{J. A. Katine\textsuperscript{2}}
\author{R. J. Hicken\textsuperscript{1}}
\author{A. P. Hibbins\textsuperscript{1}}
\author{F. Y. Ogrin\textsuperscript{1}}%
 
\affiliation{%
 \textsuperscript{1}Department of Physics and Astronomy, University of Exeter, Exeter EX4 4QL, United Kingdom
 }%
 \affiliation{%
 \textsuperscript{2}HGST, A Western Digital Company, San Jose Research Centre, CA, USA.
 }%

\date{\today}

\begin{abstract}
The curling spin wave modes of a ferromagnetic vortex confined to a microscale disc have been directly imaged in response to a microwave field excitation using time-resolved scanning Kerr microscopy.  Micromagnetic simulations have been used to explore the interaction of gyrotropic vortex core dynamics with the curling modes observed in the region of circulating in-plane magnetization.  Hybridization of the fundamental gyrotropic mode with the degenerate, lowest-frequency, azimuthal modes has previously been reported to lead to their splitting and counter propagating motion, as we observe in our spectra and measured images.  The curling nature of the modes can be ascribed to asymmetry in the static and dynamic magnetization across the disc thickness, but here we also present evidence that spiral spin waves emitted by the core can influence the spatial character of higher frequency curling modes for which hybridization is only permitted with gyrotropic modes of the same sense of azimuthal motion.  While it is challenging to identify if such modes are truly hybridized from the mode dispersion in a confined disc, our simulations reveal that spiral spin waves from the core may act as mediators of the interaction between the core dynamics and azimuthal modes, enhancing the spiral nature of the curling mode.  At higher frequency, modes with radial character only do not exhibit marked curling, but instead show evidence of interaction with spin waves generated at the edge of the disc.  The measured spatio-temporal character of the observed curling modes is accurately reproduced by our simulations, which reveal the emission of propagating short-wavelength spiral spin waves from both core and edge regions of the disc.  Our simulations suggest that the propagating modes are not inconsequential, but may play a role in the dynamic overlap required for hybridization of modes of the core and in-plane magnetised regions. These results are of importance to the fields of magnonics and spintronics that aim to utilize spin wave emission from highly localised, nanoscale regions of non-uniform magnetization, and their subsequent interaction with modes that may be supported nearby.

\end{abstract}

\pacs{Valid PACS appear here}
\maketitle


\section{\label{sec:level1}introduction}

A magnetic vortex consists of a flux-closure equilibrium state of circulating in-plane magnetization that surrounds a region of out-of-plane magnetization called the vortex core, with a diameter of only a few tens of nanometers \cite{Shinjo930}. Vortex states confined to thin film ferromagnetic discs generate negligible stray field at the edge of the disc, exhibit stability without the need for a biasing magnetic field, and can support a rich spectrum of spin waves. Vortices are therefore attractive for high density, low energy, tuneable microwave frequency components of magnetic logic, memory, and oscillator applications \cite{Magnonics,Karenowska2016,PhysRevApplied.4.047001,Chumak2014MagnonTF,PhysRevX.5.041049,articlebending,doi:10.1063/1.2089147,doi:10.1063/1.2975235}. For this reason, the magnetization dynamics of ferromagnetic elements with a vortex equilibrium state have been studied intensively, from works to acquire a greater understanding of the dynamics associated with the core \cite{doi:10.1063/1.1450816,PhysRevB.67.020403,PhysRevLett.95.167201,PhysRevB.72.024455,doi:10.1063/1.2175602,doi:10.1063/1.2738186,highgyromodes,PhysRevB.91.174425,articleWintz}, in-plane magnetized regions \cite{PhysRevLett.93.077207,doi:10.1063/1.3268453,PhysRevB.84.174401,doi:10.1063/1.4927769,PhysRevB.93.184427,PhysRevLett.122.097202} and their dynamic interaction \cite{PhysRevLett.95.167201,Guslienko_2008_split,PhysRevB.93.214437,PhysRevLett.117.037208,PhysRevB.100.214437}, to emerging signal processing applications, such as tuneable microwave emission of a spin torque vortex oscillator \cite{PhysRevB.75.140404,articlepribiag,articledussaux,Tsunegi_2014,PhysRevLett.106.167202}.
\par When an in-plane pulsed magnetic field is applied to a vortex, the lowest energy mode that can be excited is the gyration of the vortex core about an equilibrium position with uniform displacement across the magnetic film thickness, where the gyration frequency depends upon the aspect ratio of the disc \cite{YuCore,Vansteenkiste_2009}. Higher order gyrotropic modes may also be excited with nodal points in the core displacement across the film thickness \cite{highgyromodes, Guslienko2015GiantMV}. In addition, a complete set of modes related to azimuthal and radial spin waves appear \cite{PhysRevB.93.184427,azimuthalradialwaves}. Azimuthal modes exhibit a wavevector around the disc azimuth and corresponding nodal lines along its radius \cite{PhysRevLett.93.077207,PhysRevLett.95.167201}. Azimuthal modes with very high wavenumber have recently been reported in the non-linear regime \cite{PhysRevLett.122.097202}. Conversely, radial modes exhibit wavevectors along the disc radius and nodal lines of constant radius \cite{PhysRevB.84.174401}. Radial spin waves are related to Damon-Eshbach modes where their wavevector $\mathbf{k}$ is perpendicular to the equilibrium magnetization $\mathbf{M}$, since in a vortex configuration the magnetization is circulating in-plane around the core \cite{PhysRev.118.1208}.  
\par The spin wave spectrum of a vortex can be significantly different depending on the disc thickness and more generally, its aspect ratio \cite{PhysRevB.93.214437}. Spiralling spin waves found in vortex configurations have been previously explained as the hybridization of a stationary azimuthal mode and a higher order gyrotropic mode that shows no radial propagation \cite{Guslienko_2008_split,PhysRevLett.117.037208,PhysRevB.93.214437}, or alternatively as a burst of incoherent spin wave emission during a vortex core reversal \cite{Kammerer2011MagneticVC,10.3389/fphy.2015.00026}. It is well known that microscale confinement can lead to a non-uniform magnetization such as the vortex state \cite{Shinjo930}, or S and C single domain states \cite{Fruchart_2005}. Related inhomogeneity of the internal magnetic field in the region of the vortex core, or in the vicinity of edges perpendicular to an applied magnetic field, have been recently shown to be sources of spin waves due to a gradient in the magnonic refractive index \cite{PhysRevB.96.064415, article}. Such spin wave emission from these regions has been demonstrated using micromagnetic simulations and direct imaging techniques \cite{PhysRevB.96.094430, doi:10.1063/1.4995991}. 

\par In the frequency domain, techniques such as Brillouin light scattering (BLS) and vector network analyzer ferromagnetic resonance (VNA-FMR) can be used to acquire the spin wave spectra of confined nanostructures. Typically, in spatially resolved BLS microscopy (micro-BLS), the intensity of excited spin waves can be directly imaged with a spatial resolution of $<$300 nm \cite{BLS1,BLS2,BLSsquares}. A phase resolved extension of micro-BLS can also be used to quantify the spin wave amplitude through the interference of the light scattered from the spin wave with reference light modulated at the spin wave frequency \cite{doi:10.1063/1.2335627,doi:10.1063/1.3262348,doi:10.1063/1.3631756}. Magnetic resonance force microscopy of magnetization dynamics in discs can provide spatial resolution beyond the optical diffraction limit, but with limited phase information \cite{PhysRevLett.100.197601,PhysRevLett.102.177602}. On the other hand VNA-FMR can provide amplitude and phase, but no spatial information and typically averages the response of an array of magnetic elements \cite{PhysRevB.84.144406,propedges,PhysRevB.79.174433,Lara2017InformationPI}. In the time domain, time-resolved scanning transmission x-ray microscopy (TR-STXM) has been used to directly image spiral spin waves in circular discs \cite{PhysRevLett.122.117202,Behncke, articleWintz},  for which hybridization between gyrotropic modes of the core and laterally propagating, perpendicular standing spin waves was identified \cite{PhysRevLett.122.117202}. At lower frequency, time-resolved scanning Kerr microscopy (TRSKM) can be used to image the spatial character of spin waves with wavelength larger than the diffraction limited optical spatial resolution \cite{doi:10.1063/1.4995991,PhysRevB.73.134426,PhysRevB.67.020403kerrimaging,doi:10.1063/1.4998016,PhysRevB.91.174425}. 
\par In this work we report on the direct observation of curling modes with azimuthal and radial character in the in-plane magnetized region of a 40-nm thick NiFe disc with a 2-um diameter. We present micromagnetic simulations that reveal that the curling modes are excited concurrently with gyrotropic modes of the core, but also with short-wavelength, spiral spin waves that are emitted from the core. We note that the spiral spin waves emitted by the core and the curling, spiral nature of the azimuthal-radial type modes may become confused.  Hereafter, we refer to these modes as \lq spiral spin waves from the core’ and \lq curling modes’ respectively.  Our simulations allow us to explore how the spiral spin waves influence the spatial character of the curling modes, and if they play a role in the hybridization of gyrotropic modes with azimuthal and radial modes of the disc.
\par A sufficiently large disc thickness was chosen to yield a rich mode spectrum that includes azimuthal and radial modes that may coincide with higher order gyrotropic modes of the core. TRSKM with a spatial resolution of $\sim$300 nm was used to image the curling modes in the 2 um disc over a frequency range extending from 4 GHz to over 10 GHz. The simulations broadly reproduce the spatio-temporal character of the curling modes observed in the experiment.  While the core dynamics and spiral spin waves cannot be spatially resolved in the experiment, fixing the equilibrium configuration of core spins in the micromagnetic simulations prevents core dynamics and the emission of spiral spin waves, but also the azimuthal motion of the curling modes.  Fixing the core spins alters the dynamic overlap of the core and disc regions, but suggests that emission spiral spin waves may play a role in the dynamic interaction of the gyrotropic and curling modes required for hybridization.  The experimental observation of the curling therefore provides an indirect confirmation of the dynamic interaction between the core and disc regions. 

\par The disc studied in this work had a thickness of 40 nm, which has been predicted in simulations by Noske et al. to coincide with the hybridization of a counter-clockwise (CCW) azimuthal mode with the CCW first order gyrotropic mode of the core for discs with diameter of 500 nm \cite{PhysRevLett.117.037208}.  Similarly Verba et al. explained their experimental observations using simulations to demonstrate that the CCW azimuthal mode, the clockwise (CW) azimuthal mode, and the so-called \lq first curling mode', can hybridize with the CCW fundamental gyrotropic mode for all thicknesses, including 40 nm \cite{PhysRevB.93.214437}.  The evidence for hybridization was presented as the matching of the thickness dependent mode profile of the gyrotropic modes with that of azimuthal and curling modes.  The thickness dependent simulations of Noske et al. also revealed an anti-crossing at the frequency where the CCW gyrotropic and first- order azimuthal modes coincide.  Hybridization can be generally understood as the adoption of the spatio-temporal character of one mode, by another forming a completely new mode character.  This is in contrast to superposition where mode profiles overlap, sum, and lead to interference.  Here we present evidence that particular modes of our experimental observations are well aligned with the hybridized modes of vortex states in confined discs of earlier studies \cite{PhysRevLett.117.037208,PhysRevB.93.214437} and provide direct observation of the mode character predicted in simulations of Verba et al..

\section{\label{sec:level2}Methods}

\subsection{Time-resolved scanning Kerr microscopy} 

\par Time-resolved scanning Kerr microscopy (TRSKM) was used to image the spin wave modes within a single NiFe disc with diameter of 2 um and thickness 40 nm. The spin waves were imaged at remanence ($<$10 Oe) and in response to a uniform RF magnetic field applied in the plane of the disc. 
\begin{figure}[h]
\centering 
\includegraphics[trim=0cm 0cm 0cm 0cm, clip=true, width=8cm]{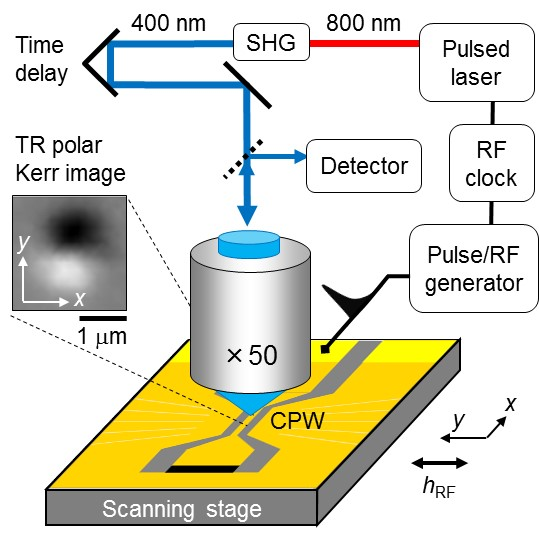}
\caption{A schematic of the time-resolved scanning Kerr microscope featuring second harmonic generation (SHG) and an optical time delay for $\sim$300 nm spatial resolution and picosecond temporal resolution.  The microscale disc was fabricated on the central conductor of a constricted section of coplanar waveguide (CPW, not to scale) where the RF field ($h_{\text{RF}}$) was enhanced, in-plane, and perpendicular to the conductor.  Images of magnetization dynamics (e.g. inset) correspond to the measured TR polar Kerr signal as the disc is scanned beneath the focused laser spot at a fixed time delay (phase) of the RF field excitation.  A $\sim$50 $\Omega$ NiCr resistor (black rectangle) was incorporated into the end of the CPW to attenuate the time-varying RF current and minimise multiple reflections.}  \label{Fields1}
\end{figure}
The RF excitation was generated using a 50 $\Omega$ impedance matched coplanar waveguide (CPW) fabricated on a sapphire substrate. The width (separation) of a short, narrow section of the CPW was 6 um (2.3 um) to maintain a characteristic impedance of 50 Ohms. The CPW and NiFe disc were fabricated from a multilayer stack of composition Ta(5)/Cu(25)/[Ta(3)/Cu(25)]\textsubscript{3}/Ta(10)/Ru(5)/ Ni\textsubscript{81}Fe\textsubscript{19}(40)/Al(1.5) (thicknesses in nm) as described in more detail elsewhere \cite{PhysRevB.91.174425}. TRSKM was carried out using a Ti:sapphire mode locked laser to generate $\sim$100 fs pulses with 800 nm wavelength at a repetition rate of 80 MHz.  Second harmonic generation was then used to generate pulses with 400 nm wavelength that were passed along a 4 ns optical time delay line, expanded by a factor of 5 and linearly polarized, before being focused to a diffraction limited spot on the surface of the disc using a high numerical aperture (NA) microscope objective lens (NA 0.6, $\times$50).  The beam was filtered to remove a residual 800 nm component and attenuated so that less than 200 uW average power was incident on the disc.  The reflected light was collected by the same objective lens so that changes in the polarization resulting from polar magneto-optical Kerr effect could be analyzed using a polarizing balanced photodiode detector.

\par Two types of measurement were performed.  First, a time-resolved (TR) scan was performed. The laser spot was positioned 0.5 um from the centre of the disc along the $+$y-direction parallel to the RF magnetic field (within the top half of the disc).  In this region the RF excitation of the in-plane equilibrium magnetization is expected to be maximum.  The polar Kerr signal was recorded while the time delay was scanned yielding a sinusoidal response corresponding to $\Delta m_{\text{z}}$ as the magnetization precesses, Fig.2(a). In the second measurement the delay was fixed at a particular time of interest, and then the disc was scanned in the xy-plane beneath the laser spot to acquire a polar Kerr image corresponding to $\Delta m_{\text{z}}$, e.g. see the inset of Fig.2(b) for polar Kerr images of the disc acquired at opposite ($+$, $-$) antinodes of the TR signal (red curve and open symbols) in Fig.2(a). For all RF frequencies used, time delays were selected so that images were acquired at similar increments in phase throughout a single RF cycle, e.g. see phases indicated by open symbols overlaid on the red curve in Fig.2(a). The resulting images were then spatially  drift-corrected and arranged according to their time delay to construct movies of a particular spin wave.  
\begin{figure}[h]
\centering 
\includegraphics[trim=0cm 0cm 0cm 0cm, clip=true, width=8cm]{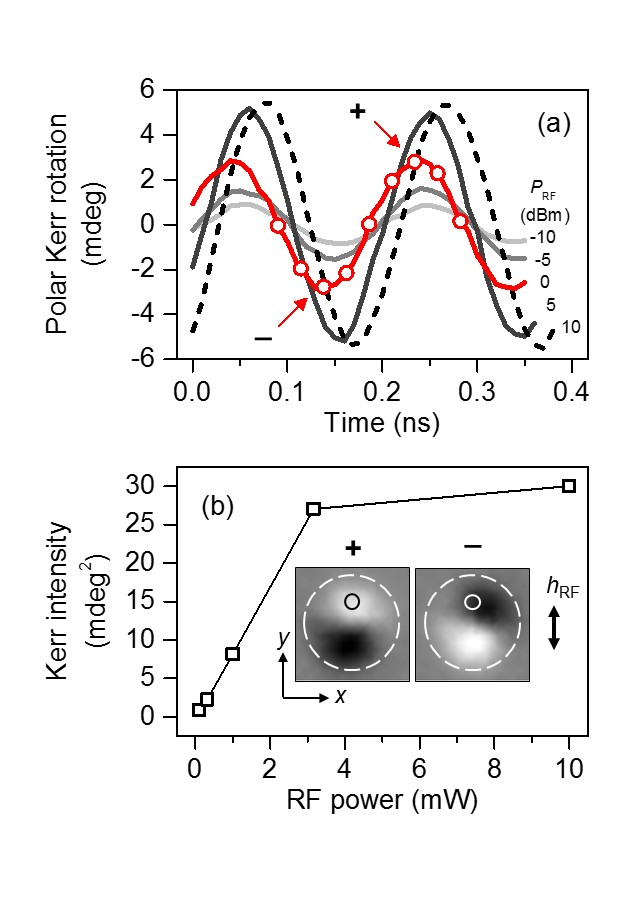}
\caption{(a) TR polar Kerr signals for an RF frequency of 5.2 GHz and RF power ($P_{\text{RF}}$) ranging from $-$10 dBm (light grey trace) to 10 dBm (black dashed trace). (b) The squared Kerr amplitude as a function of $P_{\text{RF}}$ (in mW) showing a linear dependence up to 3.3 mW (5 dBm).  Inset in (b) are polar Kerr images corresponding to the antinodes ($+,-$) of the mode excited by a RF field $h_{\text{RF}}$ with frequency 5.2 GHz and power $P_{\text{RF}}$ = 0 dBm (1 mW, red trace and symbols in (a)). The TR signals in (a) were acquired from the right hand side of the 2 um disc (large dashed circle overlaid on inset of (b)) from a small circular region corresponding to the optical spatial resolution (solid circle in inset of (b)).  For all modes imaged, the symbols on the red trace in (a) indicate the relative phase at which Kerr images were acquired.}  \label{Fields1}
\end{figure}
\par Individual spin wave modes of the vortex state were excited using RF frequencies ranging from 4.24 GHz to 10.24 GHz. The frequencies were selected from the Fourier spectrum of a TR scan acquired from the same location in the top half of the disc, but in response to a broadband pulse excitation, Fig.3(a). A pulse generator with $\sim$30 ps rise time and $\sim$70 ps duration was used to excite all modes that would couple to a uniform in-plane excitation field on picosecond timescales. The frequency and power of the excited modes was then identified from the fast Fourier transform (FFT) spectrum calculated from the TR response. To excite each mode for TR imaging the RF excitation (previously described) was applied with frequency close to that of the mode while maintaining an integer multiple of the laser repetition rate, the RF power was adjusted to compensate for the lower power of some of the modes observed in the spectrum, e.g. particularly at higher frequency, see Fig.3(b). Such modes were excited with an RF power that was enhanced by the approximate difference in power with respect to that of the highest power mode at 5.2 GHz.
\par To confirm that all modes were imaged within the linear response regime, the RF power dependence was explored for the largest amplitude mode at 5.2 GHz in the FFT spectrum, Fig.3(b). The RF power dependence on the TR polar Kerr signal is shown in Fig.2(b). At 1 mW (0 dBm) the mode excitation was within the linear regime. This suggests that the excitation of modes with lower amplitude identified in the FFT spectrum of Fig.3(b) will also be excited within the linear regime, even when the excitation power is increased to compensate for the reduced mode amplitude in the spectrum.
\par Slow phase drift of the microwave synthesiser waveform on timescales similar to that required for image acquisition can lead to mismatched spatial character of the spin wave at subsequent phases.  To minimize this, repeated TR signals were acquired from the same position in the top half of the disc ($\sim$ $+$0.5 um from center of disc) to ensure the phase (time delay) was correctly maintained.  Multiple images were acquired to ensure repeatability, and TR images were acquired in a non-sequential order to avoid a systematic accumulation of phase drift (nodal points imaged first, antinodes next, then intermediate phases to complete the series of images). Furthermore, the magneto-optical Kerr effect probes only the average response of the magnetization within the optical skin depth ($\sim$20 nm). In the simulations, variation in the phase across the thickness of the disc to a depth of 20 nm (top 5 layers of cells) was explored.  It was confirmed that the phase was approximately uniform across the disc thickness far from the vortex core.

\begin{figure}[ht]
\includegraphics[trim=0cm 0cm 0cm 0cm, clip=true, width=8cm]{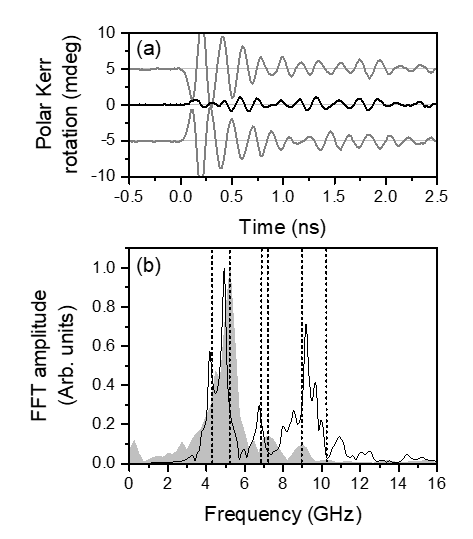}
\caption{(a) TR polar Kerr signals, excited by a 70 ps pulsed magnetic field, acquired from within the top half of the disc (top), centre of the disc (middle), and within the bottom half of the disc (bottom). (b) The fast Fourier transform spectra of the TR signal in (a) acquired from within the top half of the disc (grey shaded spectrum). Corresponding simulated spectra for the response extracted from within the top half of the disc are overlaid (solid black curve). Mode frequencies identified from the measured and simulated spectra and studied in this work are 4.24, 5.2, 6.8, 7.2, 8.96, and 10.24 GHz (vertical dashed lines).}  \label{Fields}
\end{figure}

\subsection{Micromagnetic simulations} 

To understand the observed curling, spiral nature of azimuthal and radial modes, we performed a set of micromagnetic simulations using Mumax3 \cite{doi:10.1063/1.4899186}. We simulated a disc with diameter of 2 um and thickness 40 nm with the typical material parameters of Permalloy at room temperature with saturation magnetization $\text{M}_{\text{S}}$ $= 8$ $\times$ $10^5$ Am$^{-1}$, exchange constant $\text{A}_{\text{ex}} = $ $1.1$ $\times$  $10^{-11}$ Jm$^{-1}$ and Gilbert damping constant $\alpha$ $=$ $0.008$ from a weighted average of iron and nickel \cite{doi:10.1063/1.3431384,doi:10.1063/1.4899186}. With these parameters, the single circular disc was simulated in a hexahedral grid. The grid was discretized in the x, y, z-space into 512 $\times$ 512 $\times$ 10 cells with a cell size of 3.9 nm along x and y, and 4 nm along z such that the cell size along all dimensions was smaller than the exchange length of permalloy (5.3 nm) \cite{Wang_2017}. The number of cells along x and y were chosen to be powers of $2^n$ (where $n=8$) for computational efficiency. The edges of the disc were smoothed to reduce staircase effects from hexahedral cells. The smoothed edge volume is found by averaging $p^3$ samples per cell, where $p$ is the parameter input to the function. Since the geometry is a circular disc, the \lq SmoothEdges' function was set to its maximum value ($p=8$) \cite{doi:10.1063/1.4899186}.
\par In the first stage of the micromagnetic simulations the stable equilibrium magnetization state was simulated. A vortex state with counter-clockwise circulation (circulation index 1) and core polarization towards the substrate (polarization index -1) was manually set as the initial state and then allowed to relax in a simulation with a high damping parameter ($\alpha$ $=$ 1). This particular configuration of core polarization and circulation reproduced the experimental findings as discussed in Section III. The magnetization continued to relax until  the  maximum change in induction (defined as \lq MaxTorque' parameter in Mumax3, which describes the maximum torque/$\gamma$ over all cells, where $\gamma$ is the gyromagnetic ratio of the material) reached {$10^{-7}$} T indicating convergence to the equilibrium vortex state of magnetization. The enhanced damping parameter allowed the model to relax to the equilibrium state efficiently. Once the equilibrium state was obtained the spin configuration of the disc was recorded and then used as the initial state for simulations with a pulsed magnetic field excitation. To generate a uniform  excitation over a desired frequency range, a sinc-shaped magnetic pulse was used $B_{1}(t)$,

\begin{gather}
B_{1}(t) = A_{1}\text{sinc}(2\pi f_{\text{c}} (t - t_{\text{d}})), 
\end{gather}

where {$f_{\text{c}}$} is the microwave excitation cut-off frequency, set to be 30 GHz, $t_{\text{d}}$ = 5 ns is a pulse delay and $A_{1}$ = 10 mT is the pulse amplitude. The excitation power was uniformly distributed over the selected frequency range of the pulse, so each mode up to the cut-off frequency of 30 GHz is excited with an in-plane magnetic field with amplitude of 0.3 mT. This was chosen to be sufficiently small to ensure that all modes were excited within the linear regime and to avoid any changes to the equilibrium state.  To simulate the time evolution of the spatial character of an individual spin wave mode with frequency $f_{0}$, a small amplitude continuous wave excitation $B_{2}(t)$ was applied. 

\begin{gather}
B_{2}(t) = A_{2}\text{sin}(2\pi f_{0} t). 
\end{gather}

The mode frequency $f_{0}$ was identified from the FFT spectrum of the simulated temporal response of the out-of-plane component of the magnetization $<m_{\text{z}}(t)>$, spatially averaged over a region similar in area to the focused laser spot, in response to the pulsed field $B_{1}(t)$. A sufficiently small amplitude of $A_{2}$  = 0.3 mT was chosen to ensure that each mode remains in the linear regime while driven at its resonance frequency.
A sampling period of $T_{\text{s}} = $ 25 ps was used to record 1024 simulated snap-shots of the mode spatial character, but only after the transient dynamics had subsided and the steady state was observed some time after the onset of the excitation. With these parameter values, the Nyquist criterion \cite{5055024} was satisfied for the whole range of excitation frequencies covered in this work, since the sampling frequency $f_{\text{S}}=1/T_{\text{s}}=40$ GHz is almost $\times$4 larger than the highest excitation frequency used (largest $f_{\text{RF}}= 10.24$ GHz).

\section{Results and discussion}

The TR Kerr signals acquired in response to a pulsed magnetic field, exhibit an almost identical response within the top half of the disc (0.5 $\mu$m along $+$y from the disc center) and within the bottom half (0.5 $\mu$m along $-$y from the center), but have opposite sign, Fig.3(a).  This is the expected dynamic response of regions of circulating in-plane equilibrium magnetization that lie perpendicular to the pulsed magnetic field.  Since these regions to either side of the vortex core have antiparallel magnetization, the initial torque exerted by the pulsed magnetic field will have opposite sign, leading to the observed signals in Fig. 3(a).  Clear beating of the TR signals indicates a multi-mode excitation in these regions. The average response of the core probed by the same $\sim$300 nm focused laser spot positioned at the center shows a response that has reduced amplitude and more complicated beating. The reduced net signal suggests that magnetization dynamics, such as the spiral spin waves predicted by our simulations, are detected with wavelength smaller than the focused laser spot \cite{PhysRevB.91.174425}. This leads to a smaller detectable net out-of-plane component of dynamic magnetization at the centre of the disc compared to that observed 0.5 $\mu$m to either side of the center.  
\par Comparison of the FFT spectra of the pulsed field response in the top half of the disc in the experiment (grey shaded) and in the simulation (black curve) is shown in Figure 3(b). While the complete spectral response of measured and simulated spectra show quantitative differences, overall there is good qualitative agreement of a number of the spectral peaks that have been identified (overlaid vertical dashed lines).  The differences in the amplitude of the simulated and measured spectral response, e.g. at $\sim$7 GHz and $\sim$9 GHz, is due to the uniform power delivered over all frequencies in the simulations, while the power dependence is known to be non-uniform in such experiments, e.g. see reference \cite{PaulPhysRevB.78.214412}. The main spectral peak at 5.2 GHz in the measured spectra exhibits a shoulder peak at $\sim$4.4 GHz.  This $\sim$0.8 GHz splitting is well reproduced in the simulated spectra, albeit with both modes slightly red-shifted, but within the experimental linewidth. At around 7 and 9 GHz the simulated spectrum from the centre exhibits peaks that coincide with those of the measured spectrum. The simulated peak at 6.8 GHz appears red-shifted with respect to the measured peak at 7.2 GHz, while around 9 GHz the simulated spectrum shows a number of shoulder peaks to either side of a main peak, which itself is $\sim$0.2 GHz blue shifted with respect to the measured peak at 8.96 GHz.  A weak peak in the measured spectrum at 10.24 GHz (observed at a level 20 dBm below that of the main peak on a log scale - not shown) may correspond to the simulated shoulder peak at 9.96 GHz, or to the broad higher frequency peak at $\sim$10.9 GHz, although precise identification is unclear. Modes at 5.2 GHz and 8.96 GHz  are similarly spaced  to those previously reported in a 2 um diameter disc in response to an out-of-plane RF field excitation \cite{PhysRevB.84.174401}. The modes in Ref. \cite{PhysRevB.84.174401} were shown to correspond to radial modes with 0 and 2 nodal lines along the radius of the disc excluding the center and the edge of the disc (mode n = 1 and n = 3 in Ref. \cite{PhysRevB.84.174401}).  

\par From time-resolved imaging at different frequencies we can identify the spatial character of the modes corresponding to the spectral peaks identified in the measured spectrum. It will be shown that the different curling mode character observed at different frequencies, is well reproduced by the micromagnetic simulations. The good agreement of the simulations allows the concurrent dynamics that are excited in the disc to be explored, including core dynamics, curling azimuthal and radial modes, and the possible role of that spiral spin waves play in their dynamic interaction. The discussion that follows in this section is split into two parts which address (A) a low frequency regime up to 9 GHz where we clearly observe curling mode character, and (B) a high frequency regime where a high order radial mode is observed without a strong curling character.

\subsection{Low frequency regime} 

In Fig.3(b) the most prominent spectral peaks are found in the low frequency regime for which simulations predict that spiral spin waves emitted from the core have the largest amplitude. In Fig. 4(d), 4(e), and 5(a) and 5(b), the TR spatial character of the modes is shown for frequencies (excitation power) of 5.2 GHz (0 dBm), 4.24 GHz (10 dBm), 6.8 GHz (10 dBm) and 8.96 GHz (20 dBm), respectively. Notably, Fig. 4(c) and 4(d) reveal  good agreement of the measured and simulated spatial character of the curling azimuthal mode as a function of time.

\begin{figure}[ht]
\centering 
\includegraphics[trim=0cm 0cm 0cm 0cm, clip=true, width=9cm]{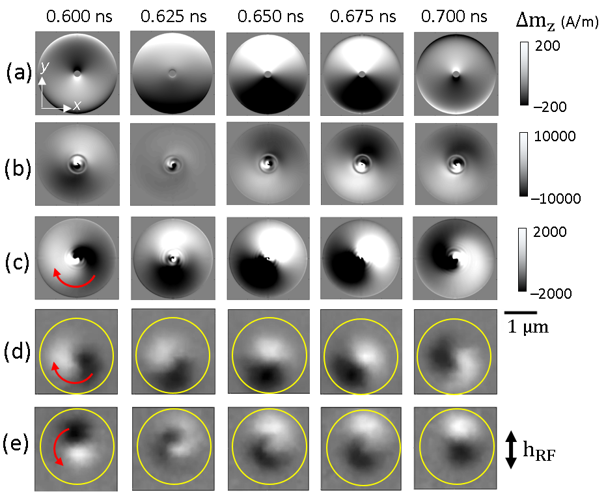}
\caption{Simulated (a, b and c) and measured (d) TR images corresponding to the out-of-plane component of the dynamic magnetization in response to an in-plane excitation of 5.2 GHz frequency. In (e) measured images are also shown for an excitation frequency of 4.24 GHz. In (a, b and c) the $m_{\text{z}}$ component is extracted from the second layer of cells from the top surface of the disc. The spins in the vicinity of the core are fixed in (a) an in a ring around it in (b) and are free to precess in (c). In (d and e) the disc perimeter is indicated by the overlaid yellow circle.}  \label{Fields3}
\end{figure}
 
\begin{figure}[h]
\centering 
\includegraphics[trim=0cm 0cm 0cm 0cm, clip=true, width=9cm]{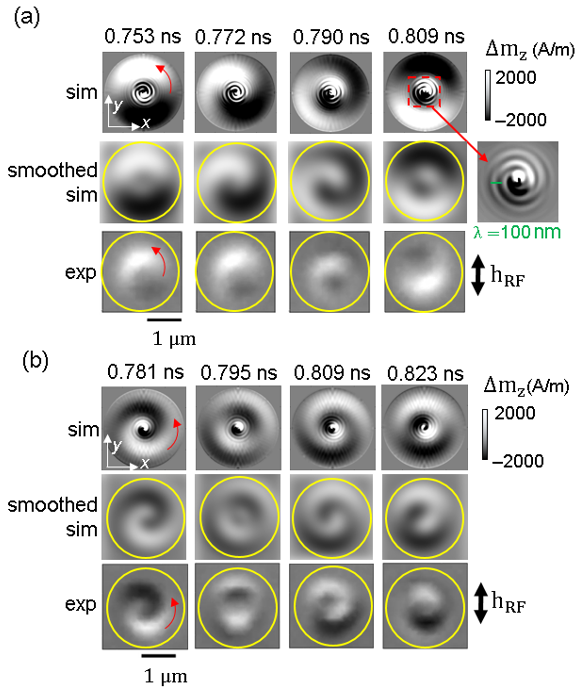}
\caption{Simulated (sim) and measured (exp) TR images corresponding to the out-of-plane component of the dynamic magnetization in response to an in-plane excitation of 6.8 GHz (a) and 8.96 GHz (b). Simulated effects of a limited spatial resolution of 300 nm are shown in the second row in (a). The $m_{\text{z}}$ component shown in the simulated images was extracted from the second layer of cells from the top surface of the disc. In (a) the inset shows the outwards propagating spiral spin wave from the core region in a larger color scale.}  \label{Fields4}
\end{figure}
\par In the low frequency regime, a whole set of azimuthal spin wave modes can arise \cite{PhysRevLett.95.167201}. At an appropriate thickness \cite{PhysRevLett.117.037208} , core dynamics have been shown to hybridize with a curling mode leading to the observed spatio-temporal mode character at certain frequencies above that of the standing azimuthal mode. Such hybridization may only be observed above a threshold thickness \cite{PhysRevB.93.214437}. To understand if the curling motion observed in our experiments is a result of dynamic coupling with the core dynamics, an equivalent disc was simulated with the spins of the core region fixed using the \lq frozenspins' function in Mumax3 \cite{doi:10.1063/1.4899186}. In this model the equilibrium spin configuration of the core, the associated dipole fields, and the exchange interaction with the rest of the disc are preserved. However, core dynamics may no longer be excited, which eliminates the dynamic interaction of the core with the rest of the disc. Fig.4(a) reveals that, when the core spins are fixed, short wavelength spiral spin waves are no longer emitted by the core and the spatial character resembles the (non curling) degenerate, lowest frequency azimuthal modes, but no longer exhibits the curling nature observed in the simulations and experiments where the core is free to gyrate (Fig.4(c) and 4(d)). To understand if the dynamic dipolar interaction between the core and rest of the disc plays a role, additional simulations were performed (Fig.4(b)) in which spins are fixed in a concentric ring-shaped region around the core. In these simulations (Fig. 4(b)) the core is free to gyrate, and spiral spin waves are emitted, but they may not propagate through the ring of fixed spins to the rest of the disc.  At the same time the fixed spins of the ring preserve the static spin configuration, dipole fields, and exchange interaction, but only allow dynamic overlap of the core dynamics with curling modes via the dynamic dipolar interaction.  Since no curling motion of the static azimuthal mode is observed, our simulations tentatively suggest that the spiral spin waves may play a role in the dynamic overlap of the core dynamics and curling modes. We note that the fixed spins models are a perturbation to the real system where all spins are free to precess. The resulting discontinuity in the dynamic dipole and exchange fields at the fixed spin boundary may also play a role in the elimination of the curling motion of the modes in the disc, but by maintaining the static spin configuration (rather than removing a ring of magnetic material, for example) influence of the discontinuity is minimized. 
\par At 5.2 GHz (Fig. 4(d)) and 4.24 GHz (Fig. 4(e)) the curling of the azimuthal mode is found to be in the opposite sense about the core. The opposite sense of azimuthal motion was reported by Guslienko et al. \cite{Guslienko_2008_split} to be the result of a dynamic dipolar hybridization of the counter-propagating azimuthal modes and the lowest frequency gyrotropic mode of the core. We note that in Ref. \cite{Guslienko_2008_split}, the azimuthal mode frequency splitting can be as large as $\sim$1 GHz, similar to our experiment (5.2 $-$ 4.24 = 0.96 GHz), and that of Ref. \cite{PhysRevLett.95.167201} for the same aspect ratio. Ref. \cite{PhysRevB.93.214437} suggests that the driven core may also hybridise with the curling mode of same sense of azimuthal motion to that of the gyration and similar mode profile across the disc thickness, which increases its frequency above the other curling mode. This is observed in our measurements, where the CW curling mode has a higher frequency than the CCW mode. Our micromagnetic simulations agree with these earlier observations by revealing the CW sense of motion of the fundamental gyrotropic mode at 4.24 GHz (Fig. 6(a) and Supplemental Material). Movies of the curling modes observed in the experiments can be found in Supplemental Material. Movies of the simulated modes and core gyrations can be found in Supplemental Material.
\par At higher frequency, hybridization of the azimuthal mode and first order gyrotropic mode may only take place when their azimuthal motion is of the same sense \cite{PhysRevLett.117.037208,PhysRevB.93.214437}. The sense of gyration, and therefore polarization, may then be inferred from TR images of azimuthal modes curling at higher frequency, such as those identified in Fig. 3(b). Ref. \cite{PhysRevLett.117.037208} suggests that a significant frequency gap can be expected in the spectrum where the first order gyrotropic mode is hybridized with the higher frequency azimuthal mode, which means that the hybridized azimuthal mode may be observed at a frequency lower than that predicted for the first-order gyrotropic mode.

\par The simulated images of Figures 4(c) (5.2 GHz), 5(a) (6.8 GHz) and 5(b) (8.96 GHz) clearly show the emission of a shorter wavelength spiral spin wave from the core. In Fig.5(a) a snapshot of the simulated core region (dashed red square) at 0.809 ns is shown in more detail (inset right), where an apparent double-arm spiral can be observed. In contrast to single-arm spirals previously reported \cite{articleWintz}, the formation of a dynamical double-dip in the core region, combined with a particular combination of the gyrotropic and the curling mode azimuthal sense of motion \cite{Kammerer2011MagneticVC}, may lead to the emission of the double-arm spiral. Spiral spin wave emission from the core was also observed in simulations at 4.24 GHz (see Supplemental Material). The spatial resolution of the experimental technique prevents the direct visualisation of these spin waves. To demonstrate this, the top row of simulated images in Fig.5(a) have been spatially down-sampled using Gaussian smoothing with a width corresponding to the optical spatial resolution of $\sim$300 nm. The smoothed simulated images (smoothed sim) are shown in the center row in Fig.5(a) and reveal greater similarity with the measured images where the short wavelength spiral spin waves emitted from the core are not resolved and lead to a reduction of the net signal in the core region. At 6.8 GHz (Fig. 5(a), 7.2 GHz (Supplemental Material), and 8.96 GHz (Fig. 5(b)) the experimental movies show an apparent inward propagation of the curling modes.  This may be understood from the curling modes at higher frequencies with superimposed azimuthal and radial components, for which the excitation of the radial component may be related to propagating spin waves from the edge of the disc towards the disc center, as discussed in Section III B.

\par The TR polar Kerr images of Figures 4(e) (4.24 GHz), 5(a) (6.8 GHz), and 5(b) (8.96 GHz) all exhibit a curling motion with the same counter-clockwise (CCW) sense, but with varying degrees of the spiral nature. This is due to a radial contribution with n = 1 and n = 3 to the mode profile at the higher frequencies of 6.8 GHz and 8.96 GHz respectively. 

\par Ref. \cite{PhysRevLett.117.037208} suggests that the curling spatial character can be more marked when the azimuthal mode is hybridized with the first-order gyrotropic mode at higher frequency. In this work at 8.96 GHz, the curling mode exhibits the strongest spiral spatial character observed at any of the studied frequencies (Fig. 5(b)) due to its high radial number (n = 3).  Furthermore,  the 8.96 GHz mode is observed at a frequency within 2 GHz of the first-order gyrotropic mode frequency. For the dimensions (2000 nm $\times$ 40 nm) and material parameters of the disc, the analytical dispersion relation from \cite{highgyromodes} yields an eigenfrequency of 10.69 GHz for the first order gyrotropic mode (n = 1), which is also expected to show CW gyration from Ref. \cite{PhysRevLett.117.037208}. When it is considered that the linewidth of the mode at 8.96 GHz is $\sim$1 GHz, that a sizable frequency gap ($>$ 1 GHz) opens in the spectrum as a result of hybridization \cite{PhysRevLett.117.037208}, and that the frequencies of the azimuthal and gyrotropic modes only need be close to hybridize, the curling mode observed at 8.96 GHz may be interpreted as a hybridized mode of the first-order gyrotropic mode and the curling mode with higher order radial component. This idea is also supported by our simulations which reveal the excitation of the first order gyromode profile at 8.96 GHz (Fig. 6(b)).  Previous work \cite{PhysRevB.93.214437} has also shown asymmetric dynamic magnetization of the first order gyrotropic mode combined with a uniform profile in the core region at frequencies as low as 6.8 GHz in discs of the same thickness (40 nm) \cite{PhysRevB.93.214437}. This was due to a superposition of the lowest order and first order gyrotropic modes hybridizing with a curling mode.  However, our micromagnetic simulations reveal that while the CCW curling modes at 6.8 GHz and 8.96 GHz agree with the experimental observations (see red arrow in Fig. 5(b)), the simulated gyrotropic motion is CW, Figure 6(d). The spiral spin wave motion exhibits a CW sense
at 6.8 GHz (see inset in Fig. 5(a)), coherently matching the sense of the 1st order gyromode. This indicates the CW gyrotropic motion and demonstrates that it is opposite to the observed motion of the curling mode. 
\par According to Ref. \cite{PhysRevB.93.214437}, CW and CCW modes are orthogonal and cannot hybridise, but the hybridization of a physical CCW curling mode may be considered via the complex conjugate amplitude of the gyrotropic mode with negative frequency and CCW motion.  It is beyond the scope of this manuscript to confirm this tentative explanation based on earlier studies, but our experimental observations nonetheless confirm the spatial character of the curling modes predicted in the simulations of Refs. \cite{PhysRevLett.117.037208,PhysRevB.93.214437} in which hybridization of curling modes with gyrotropic modes of different order were identified.

\begin{figure}[t]
\centering 
\includegraphics[trim=0cm 0cm 0cm 0cm, clip=true, width=9cm]{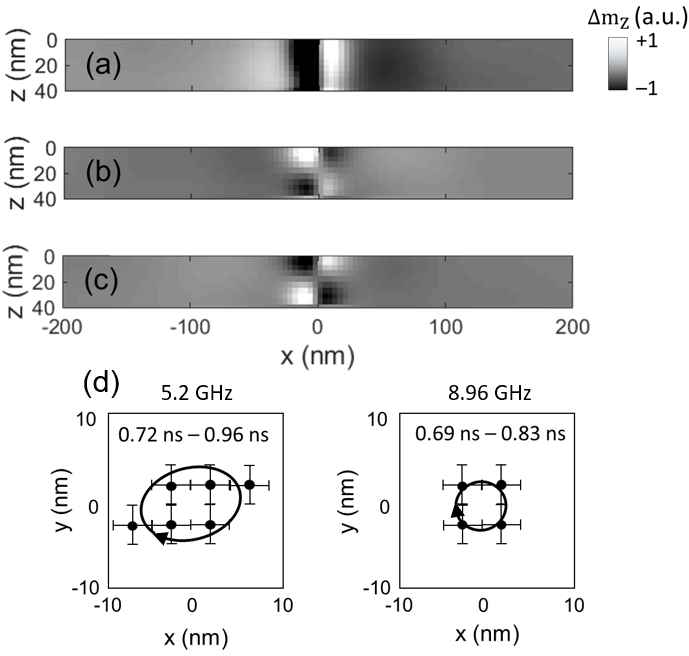}
\caption{Simulated TR images of a cross section across the thickness of the disc passing through the core region. The contrast corresponds to the normalised out-of-plane component of the dynamic magnetization in response to an in-plane excitation of 4.24 GHz (a), 8.96 GHz (b) and 10.24 GHz (c). The characteristic profile of the fundamental gyromode and that of the first higher order gyromode can be easily identified in (a), and (b) and (c), respectively. The vortex core equilibrium position is centered at x = 0 nm. The dynamic core profile of a x-z cross section obtained at 5.2 GHz is similar to that shown in (a) while the profile at 6.8 GHz is also similar to that shown in (b). (d) Simulated vortex core positions relative to the equilibrium position at (0, 0) nm, 10 nm from the top surface (black dots) for a time interval approximately equal to a period of an excitation frequency of 5.2 GHz (left) and 8.96 GHz (right). Error bars length is equal to the cell-size of the model (~3.9 nm). Black arrows are guides to the eye showing an approximate trajectory of the simulated core motion in the x-y plane (see Supplemental Material).}  \label{Fields4}
\end{figure}

\par To understand any influence of the spiral spin waves on the curling modes, radial profiles of the simulated out-of-plane component of the dynamic magnetization were extracted for the simulated mode at 6.8 GHz and are shown in Fig. 7 as a function of time. The profiles in Fig. 7(a) and 7(b) were extracted from the middle layer of cells along the  y-  and  x-direction respectively. The temporal evolution of the radial profiles reveals the onset of the spiral spin wave emission from the core and the curling motion of the azimuthal mode. Immediately after the onset of the RF field excitation, the dynamics in the disc far from the core show maximum amplitude along the y-direction where the in-plane equilibrium magnetization is perpendicular to the RF field. The initial standing nature of the azimuthal mode is confirmed by the absence of its oscillation along the  orthogonal x-direction during the first RF field cycle ($<$150 ps), and beyond a radius of 300 nm from the core where the azimuthal mode is expected. In the first two cycles of the RF excitation ($<$300 ps) this standing azimuthal mode exhibits almost constant amplitude across most of the disc radius in the y-direction where the equilibrium magnetization lies in-plane and orthogonal to the RF field.  At the same time, and within 50 nm of the center of the disc, core dynamics with a phase difference of approximately $\pi/2$ radians with respect to the azimuthal mode can be seen. These core dynamics act as the source of the radially propagating short wavelength spiral spin wave. 

\par The emission of a short wavelength spiral spin wave from the vortex core is predicted by micromagnetic simulations over the frequency range explored experimentally. Propagating spiral spin waves are coherently emitted from a gradient in the internal field close to the core, which perturbs the core from its equilibrium position \cite{articleWintz,PhysRevB.100.214437}. The emission of the spiral and the curling of the azimuthal mode is observed concurrently, while the wavefront of the emitted spiral spin wave propagates into the region of in-plane circulating magnetisation region of the disc. This occurs after almost 2.5 cycles of the RF field ($\sim$360 ps, see dashed red line in Fig. 7(a) and 7(b)) and coincides with the established curling motion of the azimuthal mode revealed as oscillations as a function of time along the x-direction with constant amplitude and phase over almost the entire radius of the disc. At larger time delay the gentle curvature of the white and black contrast indicates that at a particular time the contrast will slowly change from white to black as a function of spatial coordinate.  This is most clearly seen in Fig. 7(a) from $\sim$1.5 ns and between 0.2 to 0.6 $\mu$m. This contrast shift is due to the spiral nature of the curling azimuthal mode in the TR images and may be thought of as a time-delayed dragging of the azimuthal wavefront by the exchange interaction with the propagating spiral spin wave from the core.
\begin{figure*}[ht]
\center 
\includegraphics[trim=0cm 0cm 0cm 0cm, clip=true, width=\textwidth]{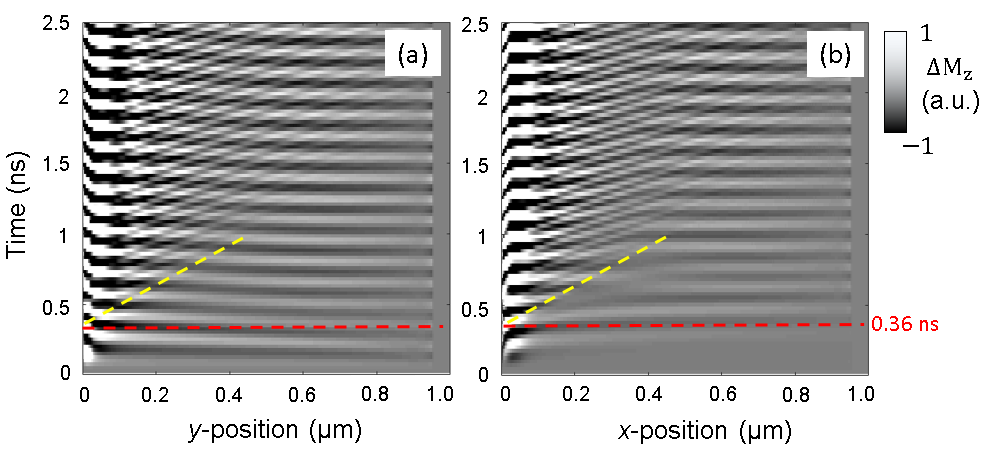}
\caption{The temporal evolution of the magnetization dynamics is shown along the radial (a) y- and (b) x-directions from the centre of the disc to its edge. The contrast corresponds to the out-of-plane component of the magnetization dynamics in response to the in-plane RF magnetic field with frequency of 6.8 GHz applied along the x-direction. The red dashed line highlights the time at which the curling makes one quarter of an azimuthal cycle and the yellow dashed line the wavefront of the propagating spiral spin wave emitted from the core.}  \label{Fields5}
\end{figure*}
\par While the onset of the spiral spin wave emission and the curling of the azimuthal mode takes place concurrently, it is interesting to identify any influence, or interaction, of the spiral spin waves on the curling mode. The short spatial wavelength ripple of the spiral spin wave contrast on the slowly varying contrast of the curling mode shows that the modes superimpose.  However, since the modes are close in frequency and occupy the same region of space and time, dynamic overlap is expected and so it might be expected that the mode character of the spiral spin waves may be adopted by the curling mode.  From Fig. 7(a) and 7(b) it can be seen that the overlap of the spiral spin wave with the curling mode reduces the distance over which the latter mode curls through quarter of the disc azimuth. For example at 1 ns, the phase of the contrast along the x-direction changes through $\pi$ radians by approximately 500 nm, in contrast to 360 ps when the same contrast shift takes place over almost the entire disc radius. Therefore, the spiral spin waves emitted by the core, may not simply be a concurrent excitation, but may also play a role in the formation of the spiral spatial character of the curling mode.  Indeed, it has already been shown in fixed spin simulations that when the spiral spin waves are not emitted from a frozen core, or prevented from propagating away from the core by a frozen ring, no curling motion of the azimuthal mode is observed.
\par In Figs. 7(a) and 7(b), the spiral spin wave propagates away from the core as time progresses, and appears as a diagonal propagation wavefront in the space-time plots of Fig. 7 (see yellow dashed line in Fig. 7(a) and 7(b)). Marked changes in the contrast as a function of x can be observed for x between 100 nm and the spiral spin wave propagating wavefront (dashed yellow line). Beyond this wavefront the contrast is almost constant, which reveals how the spiral spin wave, emitted from the core, and interacts with the curling  azimuthal mode and influences its spatial character. The wavelength of the emitted spiral spin wave at 6.8 GHz is $\sim$100 nm (see Fig. 5(a)), corresponding to a k-vector of 0.06 rad nm$^{-1}$. Experimentally it is not possible to resolve spin waves with a half-wavelength shorter than the spatial resolution of 300 nm. Furthermore, as this spin wave propagates between 0.2 $\mu$m and 0.5 $\mu$m from the core, a phase mismatch occurs between the emitted spin wave and the curling motion of the azimuthal mode, see Fig. 7(b). This leads to an apparent reduction in the amplitude of the combined dynamics within $\sim$300 nm of the core where the spiral spin waves exhibit their largest amplitude, but are averaged to a weak net signal by the limited spatial resolution of the laser spot. Consequently, these dynamics are not spatially resolved in the experiments, which can be confirmed by applying Gaussian smoothing with a full-width-half-maximum of 300 nm to the simulated images, Fig. 5(a).

\par The measured TR images of the curling azimuthal mode, and the good agreement with micromagnetic simulations, allows us to infer that short wavelength spiral spin waves are concurrently emitted from the core, propagate radially outwards and influence the spiral character of the curling azimuthal mode observed experimentally. From our work it is not possible to determine if the spiral spin waves cause the curling motion of the core, but it is already understood that curling modes can arise from an asymmetry in a radial component of the static magnetization at the top and bottom surfaces of a disc in the vicinity of the core \cite{PhysRevB.93.214437}.  However, the evidence presented in Fig. 7 reveal that the spiral spin waves can influence the spatial character of the curling azimuthal mode, and therefore suggests their dynamic overlap, which may be a mechanism for hybridisation of curling modes with gyrotropic core dynamics that are inherently coherent with the spiral spin waves. It should be noted that the dynamics presented in Fig. 7 can only be observed in the simulations since the TR images are acquired by integrating the polar Kerr signal at each pixel for at least 1 s while the dynamics in the disc are driven through more than $10^9$ cycles in that time. Therefore, the transient dynamics (spiral spin waves) that establish the spatial character steady dynamic state (the spiral nature of the curling azimuthal mode) cannot be explored with the stroboscopic TRSKM technique. 

\subsection{High frequency regime}
\par Previous studies have demonstrated that in addition to the core, other nanoscale regions of inhomogeneity of the equilibirum magnetic state can act as sources of high frequency spin waves \cite{PhysRevB.96.094430}. In this work, a higher frequency excitation of 10.24 GHz (20 dBm), revealed magnetization dynamics that extend to the very edge of the disc. In contrast, at the lower frequencies already discussed, there is a diminution of the Kerr signal in the vicinity of the disc perimeter.  This suggests that at the higher frequency of 10.24 GHz, the edge of the disc is a more efficient source of spin waves, in addition to the core. 
\par Micromagnetic simulations shown in Fig. 8 predict that at 10.24 GHz an antisymmetric radial mode is excited. The asymmetry is due to the opposite torque acting on the antiparallel in-plane equilibrium magnetization to either side of the core.  The spatial character therefore appears as the superposition of a high order radial mode with 3 nodes (excluding the core and the perimeter of the disc) and an azimuthal mode with nodal line perpendicular to the RF field and passing through the centre of the disc \cite{azimuthalradialwaves}.  This character is most clearly seen in Fig. 8(a) at 1.1712 ns and 1.22 ns when the spins of the core are fixed. When the spins of the core are free to precess, the radial-azimuthal mode appears to propagate from the edge of the disc \cite{PhysRevB.96.094430} towards the center exhibiting a spiral character that curls about the centre of the disc. The TR simulated images reveal an apparent reverse in chirality (chirality indexes $+1$ or $-1$) of the spiral pattern due to the asymmetry of the radial mode, e.g. compare simulated images in Fig. 8(b) at 1.1468 ns and 1.1712 ns.
\par In a similar mechanism to that in which the spiral spin waves appear to influence the spiral spatial character of the curling azimuthal mode at 6.8 GHz (Fig. 7), the excitation of the first order gyrotropic mode may also lead to an interaction with the higher order radial mode. A tentative explanation for the TR spatial character is the time-delayed dragging of the radial mode wavefronts by the core dynamics and resulting in a spiral wave in the immediate vicinity of the core. The propagating spiral  spin wave interacts with subsequent wavefronts at increasing time delay, and with reduced coupling, far from the dynamical core. It may be expected that a hybridization of the first order gyrotropic mode may take place with the higher order radial mode since the frequency of the gyrotropic mode, identified from Ref.\cite{highgyromodes} for the disc dimensions of this work (10.69 GHz), is sufficiently close to that of the radial mode (10.24 GHz) for dynamic overlap of the mode profiles. While it is not possible to determine if the higher order radial mode is hybridized, its low spectral power is consistent with that of a radial mode with the same order when hybridized with the first order gyrotropic mode in Ref. \cite{PhysRevB.93.214437}.  In Ref. \cite{PhysRevB.93.214437} hybridization of the higher order radial mode of a thick disc leads to a nodal line in the dynamic magnetization when the azimuthal sense of the first order gyrotropic mode is the same as the curling motion of the radial mode. The nodal line leads to a reduced net component of the dynamic magnetization and hence weaker coupling to the uniform RF field. The curling spatial character of radial modes in thick discs, such as the 40 nm disc of our study, is not necessarily a result of hybridization, but instead due to asymmetry in the radial component of the static magnetization at the top and bottom surfaces of the disc in the vicinity of the core. However, hybridization may instead manifest itself as the reduced spectral power of modes with particular sense of curling motion \cite{PhysRevB.93.214437} consistent with our observations of the radial mode at 10.24 GHz. 
\par The radial mode exhibits additional complexity whereby the tight spiral structure close to the core appears to separate to allow for the propagation of the next radial wavefront with opposite sign of $m_{\text{z}}$, but then merges with the subsequent wavefront with the same sign of $m_{\text{z}}$. The result is an apparent alternating chirality of the radial mode curling motion. At the same time, emission of spiral spin waves is also observed, but these exhibit much smaller amplitude compared to that of the radial spin waves, due to excitation with an RF frequency far from the local ferromagnetic resonance condition near the core (see animation for 10.24 GHz in Supplemental Material).

\begin{figure}
\centering 
\includegraphics[trim=0cm 0cm 0cm 0cm, clip=true, width=9cm]{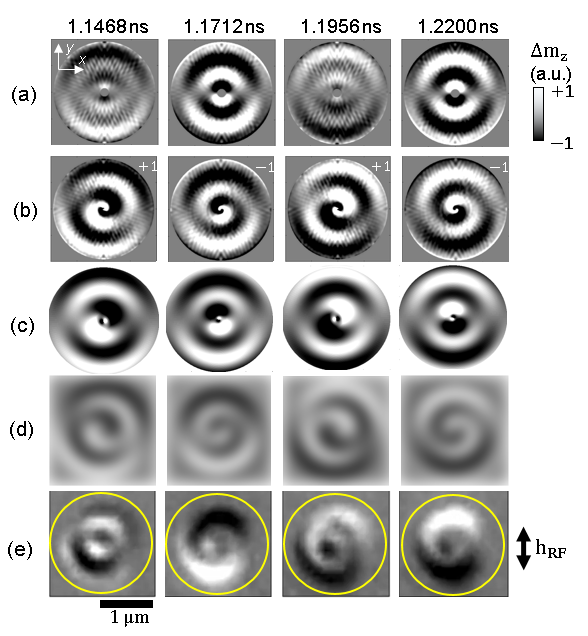}
\caption{Simulated (a, b, and d) and measured (e) TR images corresponding to the out-of-plane component of the dynamic magnetization in response to an in-plane excitation of 10.24 GHz.  In (a, b, and d) the $m_{\text{z}}$ component was extracted from the top layer of simulated cells.  The spins in the vicinity of the core are fixed in (a) and are free to precess in (b). The simulated effect of a spatial resolution of 300 nm is shown in (d).  In (c) TR images calculated from a pseudo-analytical
model from Eq.(3) and Eq.(4) and separated by T/4 ns. }  \label{Fields5}
\end{figure}

\par Fig. 9 shows the simulated amplitude of the out-of-plane component of the dynamic magnetization as a function of time extracted from three positions: approximately 70 nm (in the core region), 140 nm (at the perimeter of the core) and 210 nm (within the in-plane magnetised region) from the vortex core. The core width is approximately 300 nm, and was extracted from the simulated equilibrium vortex state (see inset of Fig. 9). The traces show that the core gyration is delayed by $\pi/4$ radians with respect to the radial wave at the perimeter of the vortex core. After $\sim$0.175 ns, the radial wave interacts with the core and the phase delay is introduced (see black dashed line in Fig. 9).

\par The dynamic out-of-plane component of magnetization of the radial spin wave ($m_{\text{r}}$) and of the core clockwise gyration ($m_{\Delta \text{g}}$) at high frequencies can be naively described in polar coordinates,

\begin{gather}
m_{\text{r}}(\rho,\theta,t) = {m}_{0}\text{sin}(\theta)\text{sin}(k_{\rho}\rho + \omega_{0} t),
\\
\begin{split}
m_{\Delta \text{g}}(\rho,\theta,t) = {m}_{0}\text{sin}(\theta - \omega_{0} t - \phi_{0}-\phi_{1})e^{-\rho/\delta_{0}}
\\-2{m}_{0}\text{sin}(\theta - \omega_{0} t - \phi_{0})e^{-\rho/\delta_{1}},
\end{split}
\end{gather}

where $\rho$ and $\theta$ are the radial and azimuthal coordinates respectively, $t$ is time, and $\omega_{0}$ is the angular frequency of the microwave excitation. The phase difference assumed between the core and the curling radial mode is $\phi_{0}=\pi/4$ (identified from Fig. 9) and $k_{\rho}$ is the radial mode wavevector. The amplitude ${m}_{0}$ is assumed to be identical in both expressions while the core dynamics are modelled as the gyration of a double-dip bipolar profile in $m_{\text{z}}$. The double dip profile has oppositely polarized regions that exist close to the moving vortex core with position described by $\delta_{0}$ ($+m_{\text{z}}$) and $\delta_{1}$ ($-m_{\text{z}}$), a phase difference $\phi_{1}$ between them \cite{Kammerer2011MagneticVC,Vansteenkiste_2009}, and with a radially decaying function since it is limited to the core region. The formation of such a double-dip may also explain the emission of the double-arm spiral spin wave as observed in Fig.5. From micromagnetic simulations of the equilibrium vortex state, the core region is estimated to be approximately $|\rho|<\delta_{0} = 0.15$  $\mu $m (see inset in Fig. 9). The outermost part of the double-dip  is delayed $\phi_{1}=\pi/8$ with respect to the innermost part of the profile ($|\rho|<\delta_{1} = 0.06$ $\mu$m) to mimic a dragging effect around the core. Assuming superposition of both waves inside the core region ($|\rho| << \delta_0$), the final pattern exhibits a spiral-like profile that appears to change chirality with time at the centre. Fig. 8(c) shows results from this pseudo-analytical model at time frames separated by $T/4$ ns, where $T$ is the period of the microwave excitation ($T = 1/f_{0}$). While this analytical model does not account for the dragging effect due to exchange interaction between spiral spin waves and the radial mode (see Supplemental Material), it does provide insight into the change in chirality of the spiral character of the simulated mode (Fig. 8(b)) as it curls around the core due to the significant phase difference between the core dynamics and the radial mode. 

Given a generic function in polar coordinates $f(\rho,\theta)$, the reverse of chirality can be described as the even symmetry $f(\theta)=f(-\theta)$. Through algebraic transformations and assuming that $|\rho| << \delta_0$ and $\phi_{1}$ is negligible, it can be trivially shown that Eq.(4) satisfies the condition $m_{\Delta \text{g}}(\theta,t)=m_{\Delta \text{g}}(-\theta,t+T/4)$, only if $\phi_{0}=\pi/4$. Together with numerical results from Fig. 8, this reveals that the phase delay between the radial mode and the core dynamics reverses chirality every $T/4$ ns.
Fig. 8(d) shows the simulated images of Fig. 8(b) after Gaussian smoothing has been applied to reproduce the 300 nm spatial resolution of the optical measurements. While the high resolution of the core dynamics in the simulated images is lost, the smoothed images yield an accurate reproduction of the measured images in Fig. 8(e). 

\begin{figure}
\centering
\includegraphics[trim=0cm 0cm 0cm 0cm, clip=true, width=9cm]{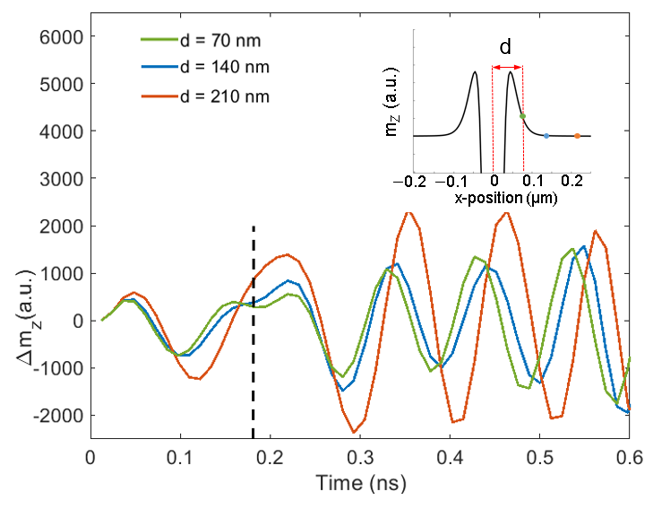}
\caption{Simulated TR traces for the dynamic out-of-plane component of magnetization extracted from the selected positions. Inset shows the core profile at t = 0 ns and the positions where magnetization is recorded as a function of time. The black dashed line highlights the time when the interaction between the radial mode and core dynamics starts.}  \label{Fields5}
\end{figure}
\par The understanding gained from the analytical model and micromagnetic simulations allows us to explore the possibility that the spatial character observed experimentally is due to a hybridization of a higher order radial spin wave in the disc with a higher order gyrotropic mode of the core. From the dispersion relation in \cite{highgyromodes}, and the size and material parameters of the disc in this work, the first-order gyrotropic mode has frequency 10.69 GHz, which lies within the line width of the observed mode at 10.24 GHz (Fig. 3(b)). Simulated profiles across the thickness and through the vortex core (Fig. 6(c)) show the characteristic first higher order gyrotropic mode profile exhibiting a single node at the centre of the disc thickness and maximum amplitude of precession close to the surfaces, but with opposite phase at the opposite surfaces at the same polar coordinate.

\par While we have explored circumstances that have been reported to be favorable for hybridization, and our measured and simulated results show consistency with those of previous studies, it remains challenging to unambiguously determine if the observed spatio-temporal character corresponds to a hybridized mode.  For the higher-order radial mode we note that core spins frozen in the equilibrium spin configuration do not lead to the curling of the radial mode observed in the experiment and simulation (Fig 8(a)). Mode curling is understood to arise from the asymmetry in the radial component of the static magnetization in the vicinity of the core at opposite surfaces of the disc, a spin configuration that is preserved in the simulations with frozen core spins.  Therefore, it seems plausible that the concurrent excitation of the spiral spin waves and core dynamics can influence the curling nature of the radial mode, as in the case of the azimuthal mode at lower frequency.

\section{Summary}
\par We have used TRSKM to image magnetization dynamics of the vortex state in a microscale disc that exhibit a curling, spiral nature. We have used micromagnetic simulations to explore the possibility that the observed mode spatio-temporal character is a consequence of hybridization of azimuthal and radial modes with gyrotropic modes of the core. Micromagnetic simulations predict the emission of short wavelength spiral spin waves from the core that cannot be observed in the measurements due to the limited spatial resolution. Micromagnetic simulations with a frozen ring of spins around the core suggest that the curling nature of the azimuthal and radial modes cannot be initiated without the emission of the short wavelength spiral spin waves from the core. The experimental observation of the curling nature in response to a microwave excitation can therefore provide indirect evidence of the emission of spiral spin waves from the core with wavelength beyond the experimental resolution. However, due to the concurrent excitation of gyrotropic core dynamics, spiral spin waves, and curling motion of azimuthal and radial modes in a thick disc, it is challenging to identify if the emission of spiral spin waves provides a dynamic coupling mechanism between the core dynamics and the curling modes to establish hybridized modes.
\par At low frequencies, the experimental TR movies of the clockwise and counter-clockwise curling of the frequency-split azimuthal modes subject to dynamic hybridization \cite{Guslienko_2008_split} with the fundamental gyrotropic mode has been observed. At higher frequency both an azimuthal and a higher-order radial mode revealed evidence of possible hybridization with the first-order gyrotropic mode of the core in accordance with previous studies \cite{PhysRevB.93.214437,PhysRevLett.117.037208}. Unlike the azimuthal modes, the higher-order radial mode also showed spin wave excitation at the edges of the disc. Micromagnetic simulations confirmed that the spin waves can propagate towards the core and play a role to establish the standing radial mode. 
\par This work provides detailed insight into the spatio-temporal character of azimuthal and radial modes of a confined magnetic vortex and the influence of gyrotropic modes of the core and propagating spiral spin waves excited concurrently. We have used an RF field excitation to continuously excite individual azimuthal and radial modes to directly observe the splitting, and curling nature of the modes predicted in simulations of earlier works \cite{Guslienko_2008_split,PhysRevB.93.214437,PhysRevLett.117.037208} and explore the possible role of the spiral spin waves on hybridization between gyrotropic core modes and the curing modes. While the frequency overlap, sense of gyration and curling, and mode thickness profile can all be considered, it remains challenging to unambiguously identify a hybridized mode from the complicated concurrent excitation of gyrotropic, curling, and propagating spiral modes, while the dispersion relations for the confined vortex provides insufficient resolution to observe the signature hybridisation anticrossings of such a rich mode spectrum.
 \par These results will permit further understanding for the control of spin wave emission from the core of a vortex and their interaction with other modes of confined magnetic elements. Such understanding will be important for the design of magnetic nanotechnologies for high frequency logic, memory and oscillator applications.

\section{Acknowledgements}
\par The authors gratefully acknowledge financial support from the UK Engineering and Physical Sciences Research Council (EPSRC) under Grant Refs. EP/P008550/1 and EP/L015331/1.

\par All data created during this research are openly available from the University of Exeter's institutional repository at https://ore.exeter.ac.uk/repository/ 

\bibliography{library}

\end{document}